\documentclass[trackchanges,linenumbers]{aastex631}

\usepackage{makecell}
\nolinenumbers

\begin{document}

\title{The NUV transit of XO-3~b}

\author[0009-0009-0960-6280]{Raven Cilley}
\affiliation{Department of Astronomy, University of Michigan, Ann Arbor, MI 48109, USA} 
\author[0000-0002-5466-3817]{L\'{i}a Corrales}
\affiliation{Department of Astronomy, University of Michigan, Ann Arbor, MI 48109, USA} 
\author[0000-0002-3641-6636]{George W. King}
\affiliation{Department of Astronomy, University of Michigan, Ann Arbor, MI 48109, USA} 

\author[0000-0002-3610-6953]{Jiayin Dong} 
\affiliation{Department of Astronomy, University of Illinois at Urbana-Champaign, Urbana, IL 61801, USA}
\affiliation{Center for Astrophysical Surveys, National Center for Supercomputing Applications, Urbana, IL 61801, USA}

\author[0000-0001-6569-3731]{Robert Frazier} 
\affiliation{Department of Astronomy, University of Michigan, Ann Arbor, MI 48109, USA} 

\author[0000-0002-5706-3497]{Kohei Miyakawa}
    \affiliation{Institute of Space and Astronautical Science, Japan Aerospace Exploration Agency, 3-1-1, Yoshinodai, Sagamihara, Kanagawa, 252-5210 Japan}

\author[0000-0002-4909-5763]{Akihiko Fukui} 
\affiliation{Komaba Institute for Science, The University of Tokyo, 3-8-1 Komaba, Meguro, Tokyo 153-8902, Japan}
\affiliation{Instituto de Astrof\'isica de Canarias, V\'ia L\'actea s/n, E-38205 La Laguna, Tenerife, Spain}

\author[0000-0003-3618-7535]{Teruyuki Hirano}
    \affiliation{Astrobiology Center, 2-21-1 Osawa, Mitaka, Tokyo 181-8588, Japan}
    \affiliation{National Astronomical Observatory of Japan, 2-21-1 Osawa, Mitaka, Tokyo 181-8588, Japan}
    \affiliation{Department of Astronomical Science, The Graduate University for Advanced Studies, SOKENDAI, 2-21-1Osawa, Mitaka, Tokyo 181-8588, Japan}

\author[0000-0002-7733-4522]{Juliette Becker}
\affiliation{Department of Astronomy, University of Wisconsin-Madison, Madison, WI 53706, USA}

\author[0000-0002-3522-5846]{James T. Sikora}
\affiliation{Lowell Observatory, 1400 W Mars Hill Road, Flagstaff, AZ 86001, USA}

\author[0000-0003-4987-6591]{Lisa Dang}
\affiliation{Waterloo Centre for Astrophysics and Department of Physics and Astronomy, University of Waterloo; Waterloo, Ontario, Canada N2L 3G1}


\begin{abstract}
\nolinenumbers

Near-UV (NUV) measurements of exoplanet transits offer a means to probe atmospheric escape, cloud formation, and planetary magnetic fields. We examine a 2024 XMM-Newton Optical Monitor NUV observation of the transit of XO-3~b, a massive hot Jupiter on an eccentric orbit with a previously observed abnormally large NUV-absorbing atmosphere. We analyze this NUV data jointly with a concurrent ground-based optical observation and all TESS transit observations, and find a NUV transit depth of $R_{p,NUV}/R_{\star} = 0.1371^{+0.016}_{-0.019}$, which is 30-70\% deeper than the optical transit. Although the optical transits do not show signs of transit timing variations, the transit center in the NUV is $22^{+13}_{-11}$ minutes late compared to the optical ephemeris. We investigate atmospheric escape as a potential explanation of the properties of this NUV transit by examining X-ray data from XMM-Newton, characterizing the X-ray luminosity of XO-3 for the first time and estimating an extremely small mass-loss rate of $\sim10^4$ \,g\,s$^{-1}$ ($\sim10^{-19}$ M$_{\text{jup}}$\,yr$^{-1}$). Finally, we investigate the likelihood of an NUV-absorbent bow-shock by estimating the magnetic field of the planet. While such a mechanism is capable of producing NUV transit offsets on the order of tens of minutes, our analytic approximations predict an early rather than late transit, indicating a need for further magnetohydrodynamic simulations.

\end{abstract}

\section{Introduction} \label{sec:intro}

The discovery of thousands of exoplanets has revealed many archetypes of planetary systems, giving us insight into the evolution of planets both inside and outside our solar system. One of the most exotic types, which make up only about 1\% of planets \citep{Wright2012}, are hot Jupiters: gas giant exoplanets with masses similar to that of Jupiter, but with short orbital periods of less than 10 days. This short and close-in orbit means they are subject to intense interactions with their host stars, including strong tidal forces and irradiation. The origins of hot Jupiters remain an open question \citep[see][for a review and references therein]{Dawson2018}. Broadly speaking, hot Jupiters are thought to form beyond the snow line in the disk of a protostar, and then migrate inwards over time \citep[e.g.,][]{Bodenheimer2000}. This migration can be gradual over millions of years, caused by friction and angular momentum exchange with the disk, or can occur rapidly due to a violent gravitational interaction with other large bodies \citep[e.g.,][]{Goldreich1980, Ford2008}. In both cases, it is expected that the orbit of a hot Jupiter will either remain circular or be rapidly circularized by tidal forces as it approaches the star. Many observed hot Jupiters have circular orbits which lie in the plane of the rotation axis of their host star. However, a fraction of observed of hot Jupiters have eccentric orbits or spin-orbit misalignments \citep[e.g.,][]{Yee2025, Albrecht2012}. These orbits may be caused by a recent disrupting interaction with another planet, and/or dynamical interactions with a companion star or outer planet via planet--planet scattering \citep{Ford2001, Beague2012, Dong2023} or secular interactions such as the von Zeipel-Lidov-Kozai mechanism \citep[e.g.,][]{Wu2003}.

One large hot Jupiter with an eccentric orbit is the planet XO-3~b, which orbits a main-sequence F5V field star in a system with no other known companions. Discovered by \citet{Johns-Krull2008}, its orbit is both eccentric and misaligned with the host star's spin axis ($e=0.2769$, $\lambda_{proj}=37^{\text{o}}$ \citep{Wong2014, Winn2009, Rusznak2025}). 
The distance to the star was uncertain until Gaia DR2, so measurements of the planet’s radius were not well constrained until fairly recently. \citet{Dang2022} analyzed two full-orbit Spitzer phase curves of the XO-3 system and reported a radius of $1.295\pm0.066$ R$_{\text{Jup}}$ and a mass of $11.79\pm0.98$ M$_{\text{Jup}}$ for XO-3~b. This study also found that, for its mass, the radius of XO-3~b in optical wavelengths may be significantly larger than expected according to planetary structure models. This could be explained by an abnormally large interior heating source equal to 20\% of the insolation. 

One explanation of the planet's inflated radius could be deuterium burning, since the mass of XO-3~b is near the deuterium burning limit. However, brown dwarf models suggest that deuterium burning ceases within the first 100 Myr of a planet’s lifetime \citep{Spiegel2011}. Another possible source of internal heat could be tidal heating, which would be logical given the eccentricity of the planet’s orbit. For tidal heating to explain the inflated radius, the orbit of XO-3~b would likely be circularizing on a timescale on the order of $\sim$Myrs \citep{Dang2022}, and we might expect to see some level of transit timing variations (TTVs). TTVs on the order of 18 minutes were detected in TESS observations by one group \citep{Yang2022}. 
However, other studies of optical transits have not found any evidence of TTVs \citep[e.g.][]{Kokori2023, Worku2022}.  

Furthermore, \textit{Swift} observations of the NUV transit of XO-3~b by \citet{Corrales2021} suggest an especially large NUV-absorbing atmosphere with an apparent radius about twice that of optical. Past studies using exoplanet transmission spectroscopy and atmospheric models point to the NUV as a tool to observe evaporated Fe{\sc ii} and Mg{\sc ii} in hot Jupiter atmospheres, probing atmospheric escape and star-planet magnetic interactions \citep{Salz2019, Sing2019, Lothringer2025}. While XO-3~b is highly irradiated by its host star, photoevaporation of the planet’s atmosphere is not expected to be efficient due to the large mass of the planet; \citet{Corrales2021} calculate a small mass-loss rate of $2\times10^{10}$ g/s. Thus, evaporated but still bound ionized atmospheric metals could contribute to the observed excess NUV absorption. 

Another possible cause of the enlarged apparent radius in the NUV could be the presence of gaseous SiO, a precursor to silicate clouds, high up in the atmosphere of XO-3~b \citep{Lothringer2020}. SiO has been considered to explain significant absorption in multiple wavelengths, including the NUV, in studies of hot Jupiters with temperatures above 2000 K \citep[e.g., ][]{Evans-Soma2025, Gapp2025, Lothringer2025, Lothringer2022}. The equilibrium temperature of XO-3~b varies throughout its orbit between $\sim$1500 K near apoapse to just above 2000~K near periapse (calculated using parameters from \citet{Dang2022} and Gaia DR2). This means silicates may condense into clouds during parts of the orbit and aerosolize or evaporate in others. With a transit approximately 15 hours after periapse, spectroscopic analysis of XO-3~b would be helpful to investigate the transitionary states of silicate clouds, hazes, and gas-phase molecules in its atmosphere, and to investigate atmospheric silicates as a cause of the observed excess NUV absorption. 



XO-3~b exhibits unusual properties that are physically poorly understood, motivating more observations to understand its formation and evolution. In this study, we analyze a new NUV observation of the transit of XO-3~b from the XMM-Newton Optical Monitor \citep[OM,][]{XMM_OM} jointly with a concurrent optical observation from an ISAS/JAXA ground-based telescope and all past Transiting Exoplanet Survey Satellite \citep[TESS,][]{TESS} transit observations. We first search for evidence of TTVs in the TESS dataset in Section \ref{sec:TTVs}, and compare our results to previous works on the topic. Then we perform a joint analysis of the TESS, NUV, and ground data to determine the radius and ephemeris of the NUV transit, in Section \ref{sec:NUV}. We discuss our results, compare them to past studies, and examine possible sources of the XO-3 system properties in Section \ref{sec:Discussion}. Finally, we conclude in Section \ref{sec:Conclusion}.

\section{TTV Investigation} \label{sec:TTVs}

\citet{Corrales2021} reported that the NUV transit of XO-3~b appeared 22 minutes early relative to the optical ephemeris, though this value was close to the ephemeris uncertainty at that time (14 minutes). TTVs, whether arising from a companion planet or orbital decay, would affect the perceived offset between an NUV transit and the predicted optical ephemeris from past orbital solutions. TTV detections for XO-3~b have been claimed in some works using transit measurements from as far back as 2008 \citep{Yang2022, Shan2023}, and are inconclusive in others which use newer TESS transit measurements or which exclude clear outliers \citep{Wang2024b, Ivshina2022}. The most recent TESS observations of XO-3~b allow for high-precision determination of the orbital solution for the modern epoch.

\begin{deluxetable}{lcccccl}[!h]
\label{tab:obslog}
\caption{Observation log}
\tablehead{\colhead{Instrument} & \colhead{Wavelength}
& \colhead{Obs Start (UTC)}    & \colhead{Obs End (UTC)} & \colhead{Exposure Time (s)} & \colhead{Transits} & \colhead{Notes}  }                                 
\startdata
XMM OM UVM2 & 200-250 nm  & 2024-02-10 0:39:31                      & 2024-02-10 13:34:34                         & 10                                    & 1                            & \makecell[l]{Fast mode; baseline correction \\ artifact;  missing $\sim$14  mins of \\ egress and post-baseline}                               \\
XMM EPIC  & 0.3-2.4 keV  & 2024-02-10 0:13:02                      & 2024-02-10 13:51:22                         & varies                                    & 1                            &  Transit is not detectable                            \\
\textit{Swift} UVOT UVM2                 & 200-250 nm  & 2019-07-14 0:10:35 & 2019-07-14 5:25:12 & varies    & 1                            &                -                         \\
\textit{Swift} UVOT UVM2                 & 200-250 nm  & 2019-10-05 00:30:31 & 2019-10-05 05:44:12 & varies    & 1                            &            -                             \\
TESS Sector 19                  & 600-1000 nm & 2019-11-28 14:08:12                     & 2019-12-23 14:27:56                         & 120                                   & 6                            &     -                                    \\
TESS Sector 53                  & 600-1000 nm & 2022-11-28 8:09:26                      & 2022-12-23 4:35:11                          & 120                                   & 6                            &     -                                    \\
TESS Sector 73                  & 600-1000 nm & 2023-12-07 7:17:14                      & 2024-01-03 3:40:26                          & 120                                   & 5                            & 1 transit cut off \\
JAXA/ISAS 1.3m        & 550-700 nm  &  2024-02-10 9:27:31 &  2024-02-10 14:06:53 &  20 & 1                            & Clouds; systematic pattern      \\
\enddata
\end{deluxetable}

\subsection{Data and Methods} \label{ssec:TTVModel}

\begin{figure}[htb!]
    \centering
    \includegraphics[scale=0.17]{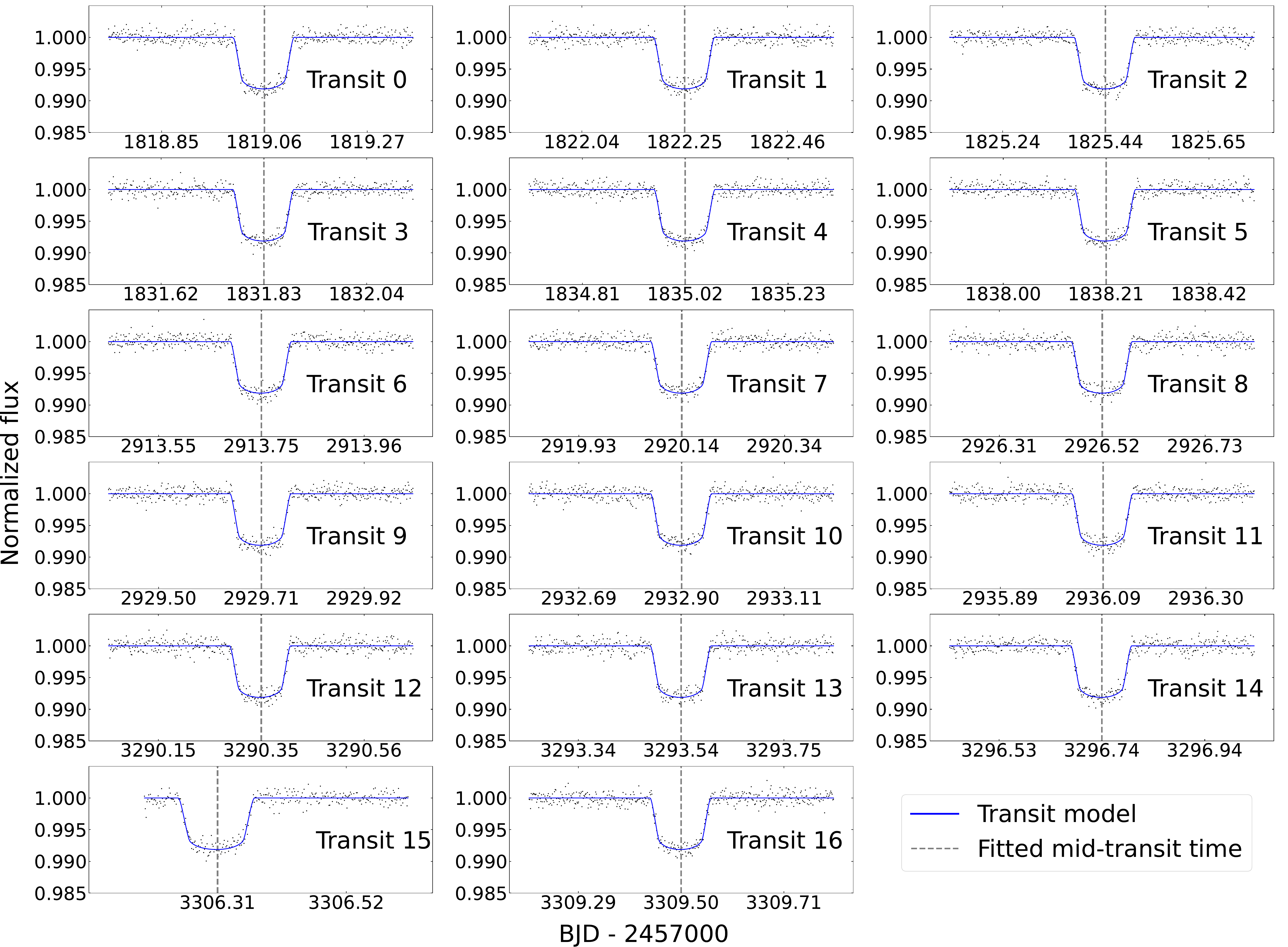}
    \caption{Detrended (GP removed) TESS light curves, with the best-fit models and mid-transit times fitted individually to each transit overplotted.}
    \label{fig:TESSResults}
\end{figure}

We investigated all previous TESS observations of XO-3~b transits to obtain an up-to-date characterization of the planet's ephemeris and to determine if transit timing variations (TTVs) were present in the optical data. We used the \texttt{lightkurve} \citep{Lightkurve} package to collect XO-3 observations at 120s cadence in TESS Sectors 19 (November 28th to December 23rd, 2019), 53 (June 13th to July 9th, 2022), and 73 (December 7th, 2023 to January 3rd, 2024). The data were downloaded from the MAST archive (DOI 10.17909/v28p-z130), and were pre-corrected and normalized using the Science Processing Operations Center (SPOC) pipeline \citep{SPOC}. 
We identified 17 transits of XO-3~b and separated them into approximately 17-hour-long light curves, keeping 7 hours of out-of-transit data before and after the approximately 3-hour-long transits. 
One observation had only 2.4 hours of pre-transit baseline measurement (Transit 15 in Figure \ref{fig:TESSResults}). We also ignored one light curve at the end of the last TESS observation which ended during the transit. We include detailed information about all datasets used in this study in Table \ref{tab:obslog}.

To fit the transits, we used the \texttt{exoplanet} package \citep{exoplanet:joss} which is powered by \texttt{pymc} \citep{Pymc} for model inference. We fit each transit with a quadratic limb-darkening transit model \citep{exoplanet:agol20} jointly with a Gaussian Process model, in order to account for correlated noise due to stellar variation and systematics in the transit light curves. The transit model free parameters were defined as follows. First, we chose uniform priors for the stellar density ($\rho_{\star}$), with a range of 0.1 to 3~g/cm$^3$. This range includes all literature estimates of the density of XO-3 \citep{Southworth2010, Stassun2017, Tsantaki2014} with ample room in the parameter space on either side. We chose a uniform prior for the ratio of the planet and stellar radii ($R_p/R_{\star}$) with a range of 0.005 to 0.5, and the impact parameter, with a range of 0 to 1$+R_p/R_{\star}$. We allowed the mid-transit times of each transit to be free parameters, rather than fitting for orbital period, in order to examine the TESS data for evidence of TTVs by manually fitting a straight line to the transit times. For each TESS mid-transit time, we used a uniform prior with a range of 2.4 hours before and after the apparent mid-transit time, chosen by eye. We used constant quadratic limb-darkening coefficients (LDCs) $u_1$=0.2126, $u_2$=0.3087 from the least-squares method catalog made by \citet{Claret2017}, with the closest matches of stellar temperature, surface gravity, and metallicity to those found for XO-3~by \citet{Stassun2017}. 
We fixed the eccentricity of the orbit to 0.2769 and an argument of periastron of 347.2 degrees, as found by \citet{Wong2014}. We use this study as a reference for eccentricity and argument of periastron because it combines transit/secondary eclipse measurements with RV measurements from \citet{Knutson2014}, providing small error bars of $0.0017$ for eccentricity and $1.7$ degrees for argument of periastron. 

To account for correlated noise in the light curves, we model a Gaussian Process (GP) for each transit using the Mat\'ern 3/2 kernel in \texttt{celerite2} \citep{celerite1, celerite2}.
The term is mathematically defined as:
\begin{equation}
    k(\tau) = \sigma^{2}\left(1 + \frac{\sqrt{3}\,\tau}{\rho}\right)
    \exp\!\left(-\frac{\sqrt{3}\,\tau}{\rho}\right),
\end{equation}
where $\sigma$ describes the amplitude of the correlated noise, and $\rho$ describes its decay timescale.
The Mat\'ern 3/2 kernel is flexible for modeling the short-timescale rough or smooth variations in the light curves caused by stellar activity or systematics.
We used a uniform prior for the unitless amplitude $\sigma$ with a range of $30\times10^{-6}$ to 1. For the variation timescale $\rho$, we used a uniform prior in the range of $1\times10^{-3}$ to $1\times10^{3}$ days. Although this range could include a stellar rotation period, we expect the model to fit to shorter-scale variability as the kernel does not encode periodic signals and we only provided the model with a few hours of out-of-transit data. We include an additional jitter term $s$ to model the flux noise.  
We then set the mean value of the GP model to the transit model, so that the modeled stellar noise would center on the transit model. 
We fit the GP and transit models jointly. Including the transit and GP parameters, the model contained 3 parameters which were shared amongst all TESS transits ($\rho_{\star}$, $b$, $R_p/R_{\star}$), and 4 parameters which were individual to each transit light curve (the mid-transit time and the 3 GP parameters).

To infer the stellar and planetary parameters, we first ran an optimization using the \texttt{find\_MAP} function in \texttt{pymc},
and then used a Hamiltonian Monte Carlo (HMC) sampler, No-U-Turn Sampler (NUTS), to find the posteriors.
We run 5000 burn-in steps and draw 3000 samples for 10 chains. 
We used a target accept rate of 0.9 to reduce divergences while maintaining computational efficiency.

\subsection{Results} \label{ssec:TTVResults}

\begin{figure}[htb!]
    \centering
    \includegraphics[scale=0.51]{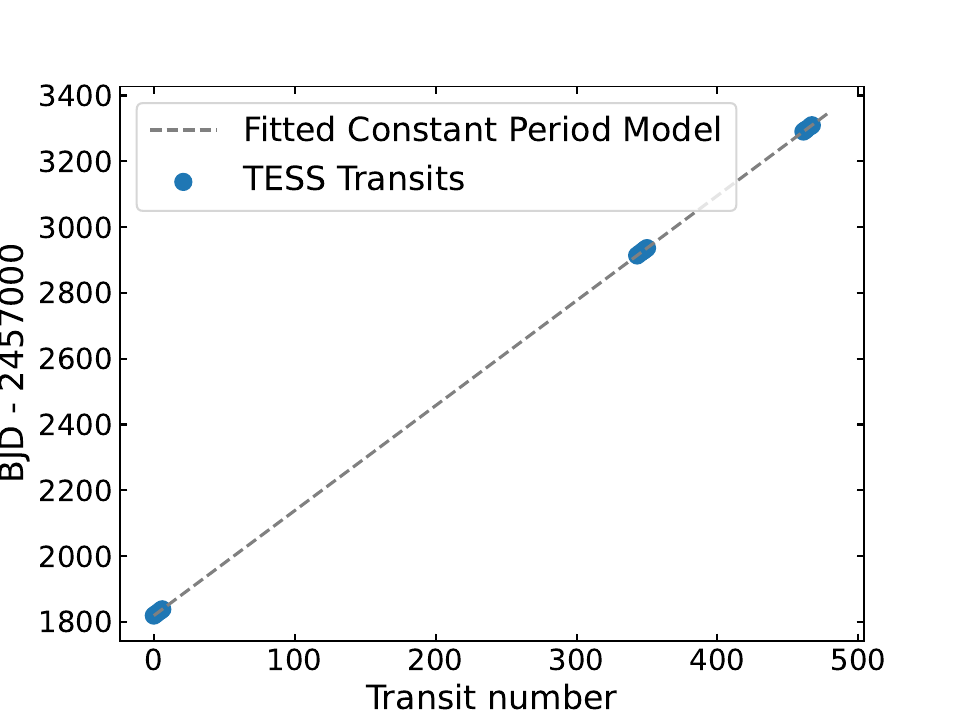}
    \includegraphics[scale=0.41]{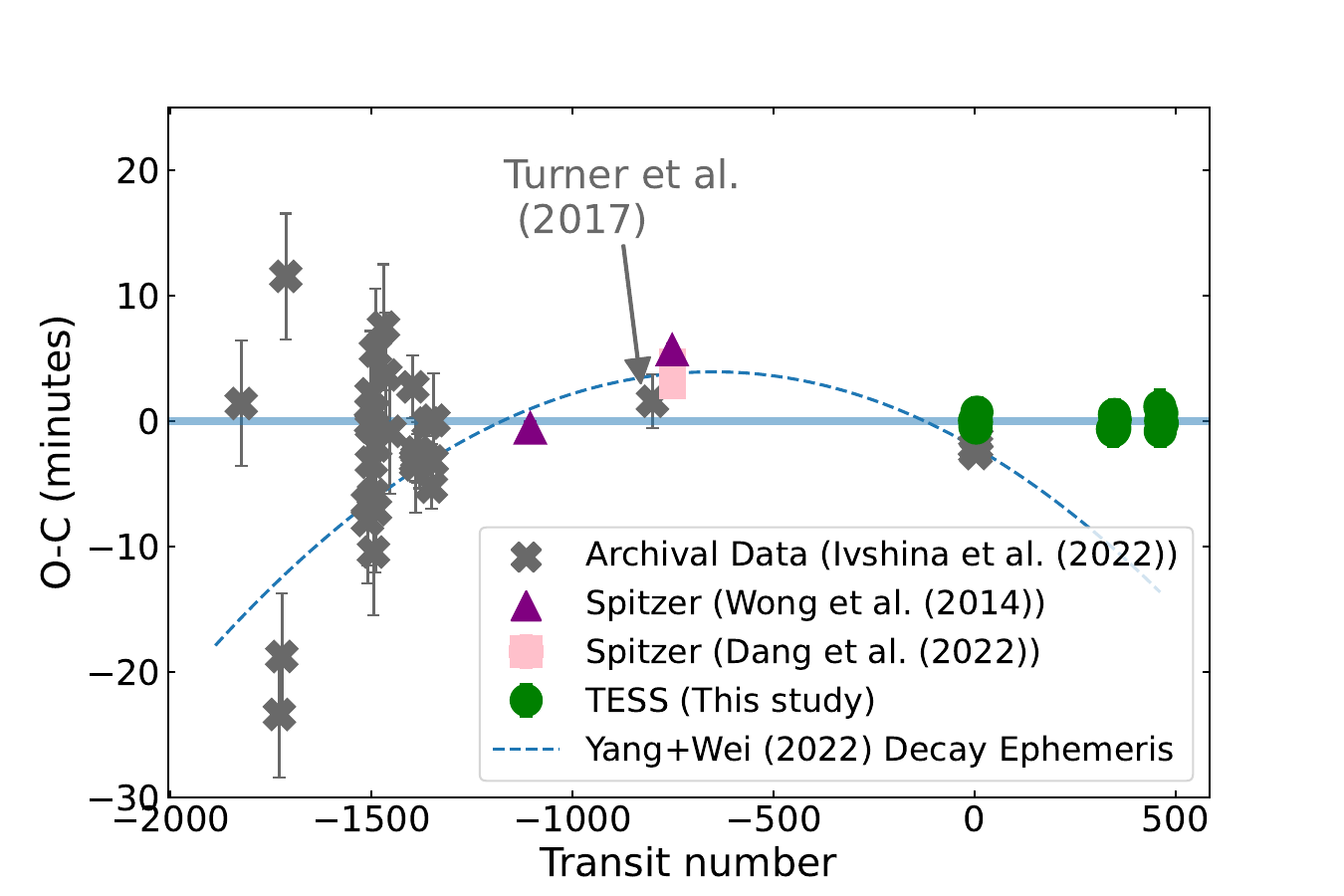}

    \caption{Transit timing analysis results. (Left) The best-fit mid-transit time for each of the TESS transits compared to the transit number are overlaid with the best-fit line (gray), where the slope gives the orbital period. (Right) The difference between the observed mid-transit time and the expected mid-transit time calculated from the best-fit orbital period from the TESS dataset. Overlaid are the Spitzer results from \citet{Wong2014} and \citet{Dang2022}, as well as optical archival data reported in \citet{Ivshina2022}. The dotted line shows the difference between our best-fit ephemeris and the period-decay ephemeris from \citet{Yang2022}. The increasing difference of the TESS mid-transit times from the decay ephemeris indicates that the TESS transits are not consistent with these authors' period-decay model. The first Spitzer data point is consistent with our ephemeris, while both studies of the second Spitzer transit find mid-transit times offset by 3-6 minutes. However, one transit analyzed by \citet{Turner2017} occurred shortly before the second Spitzer transit, and was consistent with our ephemeris.} 
    \label{fig:TESSTiming}
\end{figure}

The best-fit transit models along with detrended TESS data are shown in Figure \ref{fig:TESSResults}. Figure~\ref{fig:TESSTiming} (left) shows the transit number versus fitted mid-transit time and a linear function fit to the points. Figure~\ref{fig:TESSTiming} (right, green points) shows the residuals. These results show no evidence for TTVs over the course of five years and within the uncertainties of the TESS transit timings, which are on the order of two minutes or less. The best-fit slope for the mid-transit time points gave an orbital period of $3.1915234\pm10^{-7}$ days, which is consistent with some previous studies \citep[e.g.][]{Stassun2017}, but inconsistent past the 5th or 6th decimal place (on the order of seconds) with others \citep{Southworth2010, Bonomo2017, Kokori2023}. 

Our null TTV result is in disagreement with a previous study of XO-3~b which concluded that the 2019 TESS transits are approximately 17.6~minutes late relative to the orbital period derived by \citet{Wong2014}, indicating the presence of TTVs \citep{Yang2022}. They estimate a change in orbital period of $-6.2\times10^{-9}\pm2.9\times10^{-10}$ days per orbit per day (a change in orbital period of about 1 second per 5 years), with an orbit decay timescale of 1.4 Myr. We note that their TTV detection is strongly influenced by two data points obtained with Spitzer \citep{Wong2014, Dang2022} in the intervening years between the discovery of XO-3~b \citep{Johns-Krull2008} and today \citep{Shan2023}. We compare our TESS transit timings to the orbital decay model presented by \citet{Yang2022}, the referenced Spitzer transits, and a set of archival transits \citep{Johns-Krull2008, Winn2008, Hebrard2008, Turner2017,Ivshina2022} in Figure \ref{fig:TESSTiming} (right). The increasing deviation of the TESS mid-transit times from the ephemeris presented by \citet{Yang2022} indicates that the TESS observations are not consistent with these authors' adopted orbital decay model.

There is a significant scatter in the earliest observed mid-transit times, which do not follow either ephemeris. The first Spitzer transit analyzed by \citet{Wong2014} is consistent within 1 minute of our slope-derived ephemeris, but the second Spitzer transit analyzed by both \citet{Wong2014} and \citet{Dang2022} is between 3 and 6 minutes from the expected midpoint given by our ephemeris. This could be an indication of TTVs, following the interpretation of \citet{Yang2022}. However, we also note the transit timing result of \citet{Turner2017}, which occurred approximately 156 days before the second Spitzer transit and was consistent within 1$\sigma$ of the expected mid-transit time given our best-fit ephemeris. This indicates that, considering all transit observations, there is not strong evidence for TTVs that can be explained by orbital decay. 


\section{NUV Transit Measurement} \label{sec:NUV}

We found no evidence of TTVs in the TESS data from Nov 2019 to January 2024. This dataset provides a good constraint for the orbital parameters at the time of our XMM-Newton Optical Monitor (OM) observation of the XO-3~b transit in 2024. We turned to investigating whether the $R_p/R_{\star}$ and mid-transit time varies between the optical and NUV wavelengths by analyzing the optical data simultaneously with the XMM-Newton OM observation.

\subsection{Data} \label{ssec:NUVData}

We analyzed a 13-hour XMM-Newton OM observation of XO-3 during transit taken on February 10th, 2024 (Obs. ID 0922340301), which utilized the UVM2 filter (200-250 nm). More information about this dataset is listed in Table \ref{tab:obslog}. We reduced the XMM-Newton OM ``fast mode'' data using the Science Analysis System (SAS) version 21.0.0 and the methods described in the SAS Data Analysis Threads\footnote{\url{https://www.cosmos.esa.int/web/xmm-newton/sas-threads}}. The light curve created by the standard \texttt{omfchain} processing chain was placed into 10 separate files, corresponding to each ``image mode'' exposure. We binned each exposure to 30 points, amounting to approximately 2.5 minutes per bin. In the fourth and fifth exposures from the start of the observation, localized completely before the transit occurs, there was a significant increase in the median detected count rate. This jump also corresponds to the time in which the star has drifted the furthest out of the center of the fast mode window. Since this sudden increase in baseline count rate completely spans two exposure frames and corresponds to the location of the largest drift correction, we assumed it is an artifact of a correction done in the \texttt{omfchain} process. We accounted for this issue by normalizing the data using different baselines for the lightcurve sections inside and outside of the affected image frames. We calculated the baseline within the instrumental jump by averaging the values of all affected points. We calculated the baseline of all other points by averaging the values of points before the expected start of the transit, using the ephemeris we calculated earlier from fitted TESS mid-transit times with a buffer of 30 minutes. This process made the light curve look mostly continuous by eye. We do not expect this anomaly to affect our transit analysis, because the end of the final affected frame is separated from the expected ingress of the transit by $\approx$ 4 hours. We also note that the XMM-Newton OM observation ended near the expected end of the transit, and did not capture the post-transit baseline or possibly the full egress of the transit.

We obtained optical ground-based observations of the XO-3~b transit simultaneous with the XMM-Newton observation on 2024-02-10 (UT) using the 1.3m telescope in JAXA/ISAS near Tokyo.  The observation was interrupted by some technical trouble and occasional clouds, but we were able to obtain more than 50\% of the transit, including the ingress ($>50\%$) and egress ($>90\%$). A standard photometric pipeline was applied, and the transit was clearly detected. However, the light curve was also affected by a systematic pattern, likely due to the stellar centroid motion on the detector, substantiating the need to include a GP analysis with this dataset as well. While this data is not of sufficient quality to draw conclusions about transit timing at its epoch, when combined with the concurrent XMM-Newton OM data it helps constrain the potential offsets between optical and NUV transit.

\subsection{Joint NUV and Optical Model} \label{ssec:NUVModel}

We created a model to analyze all three sets of optical and NUV data together. We simultaneously fit the transit and GP models using the same process as stated in section \ref{sec:TTVs}. We utilized some of the same parameters as described in section \ref{ssec:TTVModel}: $b$ and $\rho_{\star}$ were shared among the three datasets in both wavelengths. Each transit light curve was assigned a separate GP model, also using the priors described in section \ref{ssec:TTVModel}. We defined two radius parameters for the system: one for optical, which applied to the TESS and ISAS data ($R_{p,{\rm opt}}/R_{\star}$), and one for the XMM-Newton NUV data ($R_{p,{\rm NUV}}/R_{\star}$). The prior for $R_{p,{\rm opt}}/R_{\star}$ was identical to the prior we used in our TTV analysis. For $R_{p,{\rm NUV}}/R_{\star}$, we defined a uniform prior with a range of 0.005 to 0.2. 

Since TTVs were not observed in the TESS data, we replaced the optical mid-transit time parameters with a fit for the orbital period of XO-3~b. This orbital period only described the transit ephemeris for the optical (TESS and ISAS) data, not the NUV data. We used a uniform prior with a $\pm5\sigma$ range around the orbital period found by \citet{Kokori2023}, given in Table~\ref{tab:results}. We used the first TESS transit in our dataset to define the reference mid-transit time for the orbital solution, using a uniform prior with the range set to the whole visible transit. 

The NUV mid-transit time was left as a free parameter, to allow us to compare the NUV transit time with the expected time derived from the optical ephemeris. We used a uniform prior, with a range of 1.2 hours before and 1.7 hours after the mid-transit time from the ephemeris prediction of \citet{Kokori2023}.  This range spans the from the visual ingress of the transit to the end of the observation. Before considering limb darkening coefficients, this joint model had 2 parameters which were shared amongst all transits ($\rho_{\star}$, $b$), 3 parameters which were only shared amongst optical transits (TESS and ISAS; orbital period, reference mid-transit time, and $R_{p,{\rm opt}}/R_{\star}$), two parameters which only applied to the NUV transit ($R_{p,{\rm NUV}}/R_{\star}$ and NUV mid-transit time), and the three GP parameters for each transit individually. 

Limb darkening coefficients (LDCs) are degenerate with the impact parameter, which affects the $R_p/R_\star$ in a transit model. Previous studies show significant discrepancies between calculated LDCs in the TESS band depending on the model used \citep{Patel&Espinoza2022, Claret2017}, so the most accurate estimate of LDCs for this star is unclear. To analyze how LDCs may affect our results, particularly the fitted $R_p/R_{\star}$ values, we fit the transit models three times. The first model had fixed optical LDCs, described in section \ref{ssec:TTVModel}, and fixed NUV LDCs, which we calculated with the Limb Darkening Toolkit \citep[LDTK,][]{ldtk} using XO-3 stellar parameters from \citet{Stassun2017} and \citet{Bonomo2017}, and the XMM-Newton OM response files for UVM2. The toolkit returned quadratic NUV LDCs $u_1$=1.553, $u_2$=-0.641. For the second fit, we introduced quadratic optical LDCs as free parameters in the model, using the \texttt{QuadLimbDark} prior distribution from \textsf{exoplanet}. This distribution is based on the \citet{Kipping2013} quadratic limb darkening law, and provides an uninformative prior. This added two parameters to the joint model, which only applied to the optical transits (TESS and ISAS). 

We conducted a third fit with both the optical and NUV LDCs as free parameters. However, we do not include these results because freeing the NUV LDCs did not improve the goodness-of-fit statistics and merely increased the uncertainty of the results. 
The best-fit NUV LDCs ($u_1=0.7^{+0.6}_{-0.5}$, $u_2=-0.05^{+0.5}_{-0.5}$) were far from the expected values given by the Limb Darkening Toolkit, and had large uncertainties. This is likely because the XMM-Newton OM light curves lack the precision to constrain NUV LDCs. We expect that the LDCs given by the Limb Darkening Toolkit are reasonable, since we used response files from the instrument and provided measured XO-3 stellar parameters from the literature. Thus, fitting for NUV LDCs introduces unnecessary error and produces arbitrary results with the insufficient quality of our NUV data. 

\subsection{Joint Analysis Results} \label{ssec:NUVResults}

\begin{deluxetable}{lcccccc}
\tablecaption{Results of simultaneous fit to TESS, ground, and XMM-OM data}
\label{tab:results}
\tablehead{\colhead{} & \colhead{Fixed LDCs} & \colhead{Free Optical LDCs} & \colhead{Prior (If Applicable)}}
\startdata
\textbf{Optical}  \\
\hline
$R_{p, {\rm opt}}/R_{\star}$ &  $0.08835\pm{0.00028}$  &$0.08868^{+0.00088}_{-0.00068}$ & $\mathcal{U}$(0.005,0.5)\\
$R_{p,{\rm opt}}$ ($R_J$) &  $1.324\pm{0.095}$  &  $1.329\pm{0.095}$ & - \\
$u$ & (0.2126, 0.3087)$^a$ & ($0.24^{+0.21}_{-0.16}$, $0.24^{+0.21}_{-0.28}$) & \texttt{QLD}$^b$\\
\hline
\textbf{NUV}   \\
\hline
$R_{p, {\rm NUV}}/R_{\star}$ &  $0.1354^{+0.016}_{-0.019}$  &  $0.1371^{+0.016}_{-0.019}$  & $\mathcal{U}$(0.005,0.2)\\
$R_{p, {\rm NUV}}$ ($R_J$) &  $2.03^{+0.27}_{-0.31}$  &  $2.05^{+0.28}_{-0.31}$ & - \\
midt (BJD-2457000)  &  $3351.0123^{+0.0087}_{-0.0075}$   &  $3351.0124^{+0.0086}_{-0.0074}$ &  $\mathcal{U}$(expected-0.05, expected+0.07)$^c$ \\
$u$ & (1.553, -0.641) & (1.553, -0.641)  &  Fixed from LDTK\\
$\Delta t_{mid}$ (minutes) &  $22.20^{+12.46}_{-10.83}$   &  $22.23^{+12.40}_{-10.76}$  & -\\
GP $\log \rho$ & $-6.19^{+0.33}_{-0.25}$ &  $-6.19^{+0.33}_{-0.25}$ & $\mathcal{U}$($\log 10^{-3}$, $\log 10^3$)\\
GP $\log \sigma$ & $-4.33^{+0.11}_{-0.16}$ & $-4.33^{+0.11}_{-0.16}$ &$\mathcal{U}$($\log 10^{-3}$, $\log 10^3$)\\
\hline
\textbf{System}    \\
\hline
Eccentricity ($e$) &  $0.2769^d$  & $0.2769^d$ & -- \\
Argument of Periastron ($\omega$, deg) &  $347.2^{d}$  & $347.2^{d}$ & -- \\
Impact parameter ($b$) &  $0.689\pm{0.010}$  & $0.700^{+0.013}_{-0.014}$ & $\mathcal{U}$(0.1, 1+$R_p/R_{\star}$) \\
Orbital period (days) & $3.1915234\pm(5\times10^{-7})$  & $3.1915234\pm(5\times10^{-7})$ &     $\mathcal{U}$(3.191525$\pm0.0007$)\\
Stellar density ($\rho_{\star}$) (g/cm$^3$) &  $0.687\pm{0.027}$  &  $0.669^{+0.031}_{-0.029}$ & $\mathcal{U}$(0.1, 3)\\
\hline
\multicolumn{7}{l}{$^a$Fixed to values from \citet{Claret2017}} \\
\multicolumn{7}{l}{$^b$\texttt{xo.distributions.QuadLimbDark} (\texttt{QLD}) is an \textsf{exoplanet} class, providing a distribution of quadratic} \\
\multicolumn{7}{l}{limb darkening coefficients for transiting planets.}  \\
\multicolumn{7}{l}{$^c$Expected value based on ephemeris from \citet{Kokori2022}. The asymmetry in the mid-transit time prior accounts} \\
\multicolumn{7}{l}{for the ending portion of the transit that may not have been captured in the XMM-Newton observation.}\\
\multicolumn{7}{l}{$^d$Fixed from \citet{Wong2014}} \\
\enddata
\tablecomments{Values with error bars give the median of the posterior results and the $1\sigma$ credible interval. For priors, the $\mathcal{U}$ symbol indicates a uniform distribution with the lower and upper limits given.}
\end{deluxetable}

\begin{figure}[htb!]
    \centering
    \includegraphics[scale=0.25]{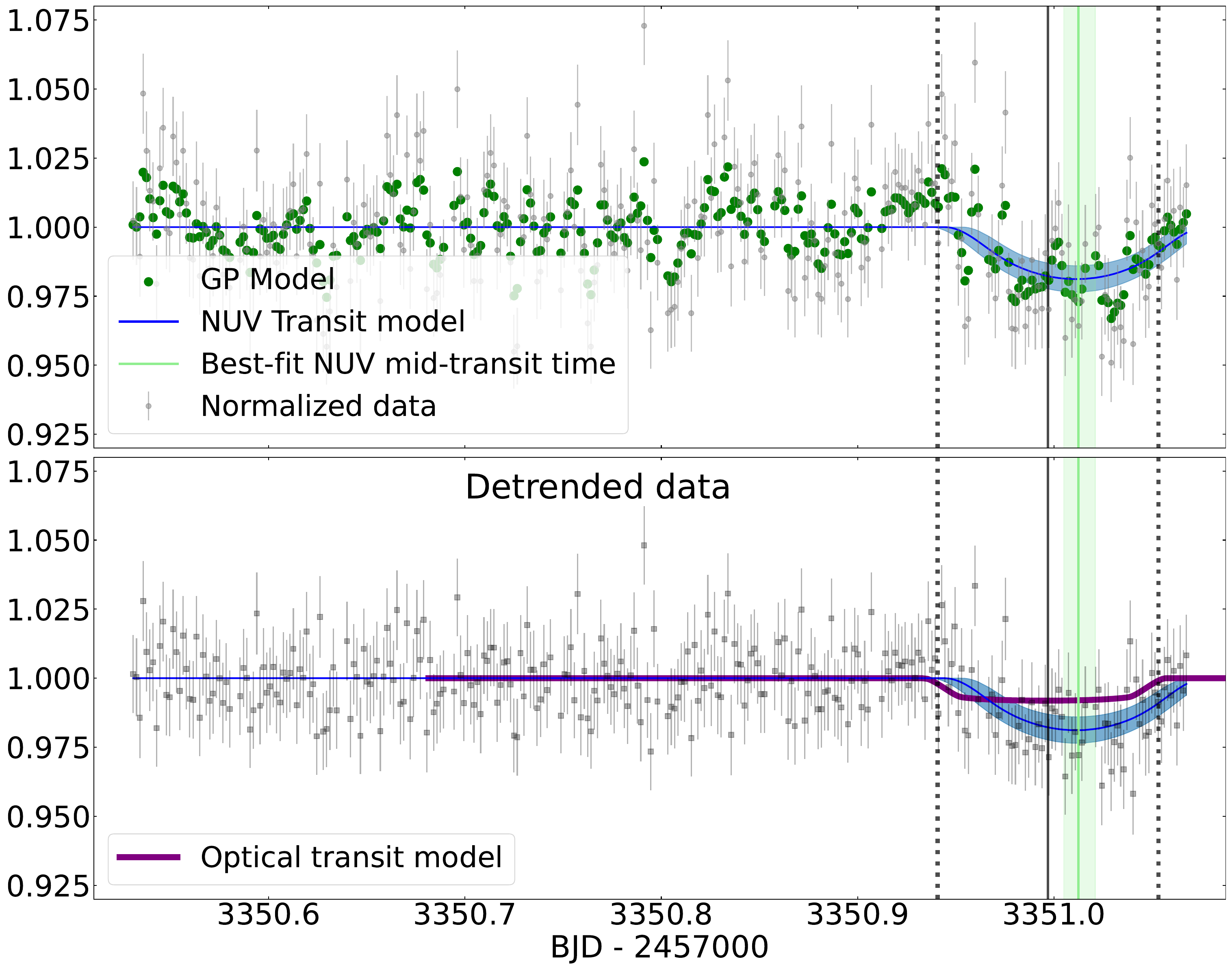}
    \caption{Results of our analysis of XMM-Newton OM data. The top panel shows the normalized data and expected duration of the NUV transit, using our ephemeris fitted to optical TESS data. The bottom panel shows the detrended data, which is equal to the normalized data divided by the GP model. The bottom panel additionally provides a comparison between a forward-propagated optical transit model and the NUV transit model.}
    \label{fig:XMMResults}
\end{figure}

Table~\ref{tab:results} shows our best-fit results with the optical LDCs fixed (column 1) versus free (column 2), along with the priors for each parameter. The Free Optical LDCs  model (column 2) 
returned best-fit parameters, including LDCs, that were similar to the results from the more constrained model (Fixed LDCs, column 1) without significantly increasing the uncertainties. 
Thus, in the following discussion, we focus on the results from the Free Optical LDCs model. Figure \ref{fig:XMMResults} shows the detrended XMM-Newton OM light curve and the best-fit transit model.

The best-fit orbital period was identical within 1$\sigma$ to the orbital period we calculated during the TTV investigation (Section~\ref{sec:TTVs}). The best-fit $R_{p,{\rm opt}}/R_{\star}$ corresponded to a radius of $1.329 \pm 0.095 R_J$ using the $R_{\star}$ value of $1.54\pm 0.11 R_\odot$ from \citet{Stassun2017} using Gaia DR2 \citep{Gaia2016, GaiaDR2}. This is within $1\sigma$ of the $R_p/R_\star$ values from \citet{Wong2014} and \citet{Kokori2023}. The best-fit $R_{p,{\rm NUV}}/R_{\star}$ of $0.1371\pm0.02$ corresponded to a radius of $2.05\pm0.3R_{\rm J}$, which is ~$\sim55\%$ larger than optical radius. Interestingly, the XMM-Newton OM best-fit mid-transit time was $22^{+13}_{-11}$  minutes late compared to our best-fit optical ephemeris, but still consistent with the optical value within 2$\sigma$. We discuss the potential implications of this measurement in Section~\ref{sec:NUVtiming}. We also note that the best-fit transit model includes $\sim$14 minutes of egress after the end of the XMM-Newton observation. 

\begin{figure}[htb!]
    \centering
    \includegraphics[scale=0.54]{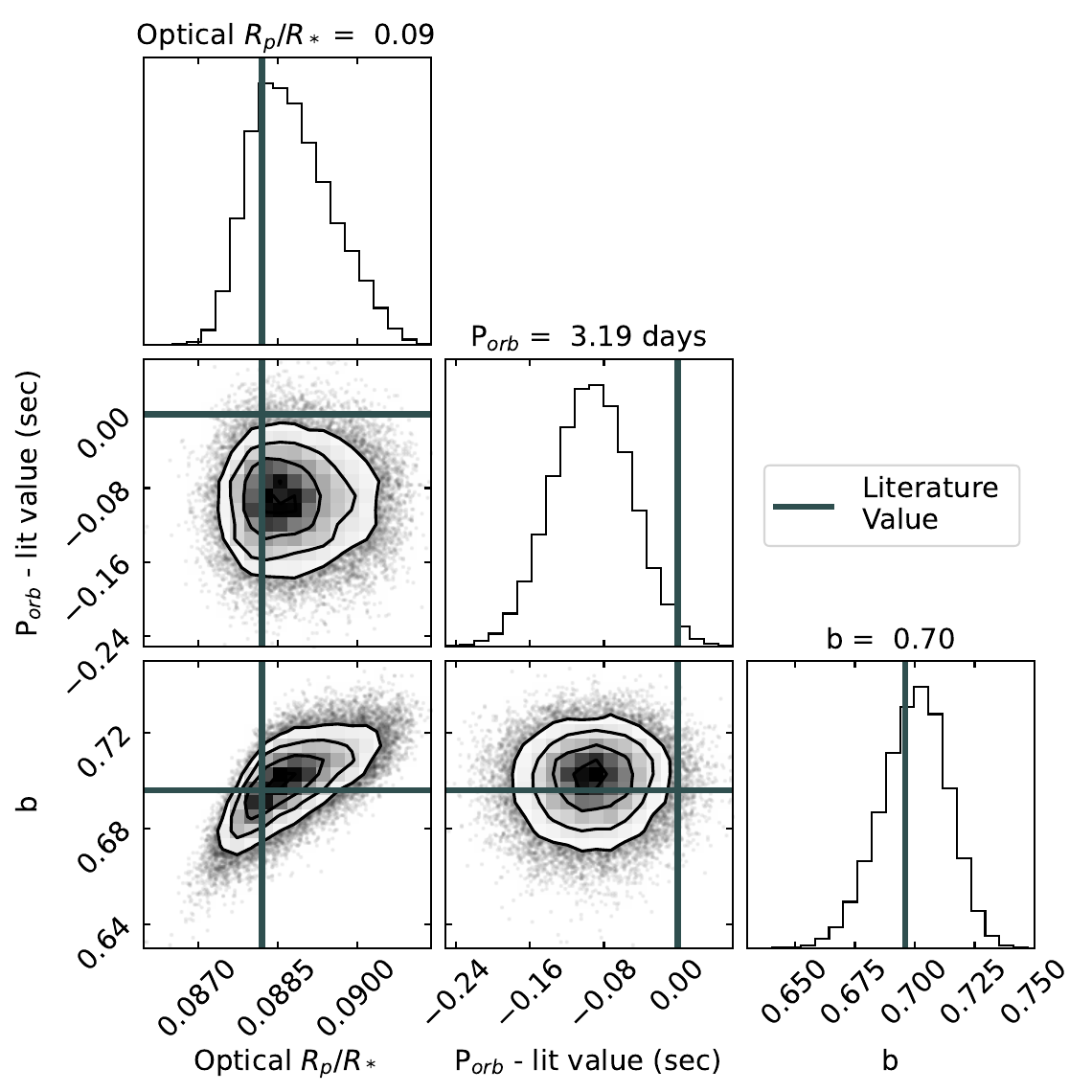}
    \includegraphics[scale=0.54]{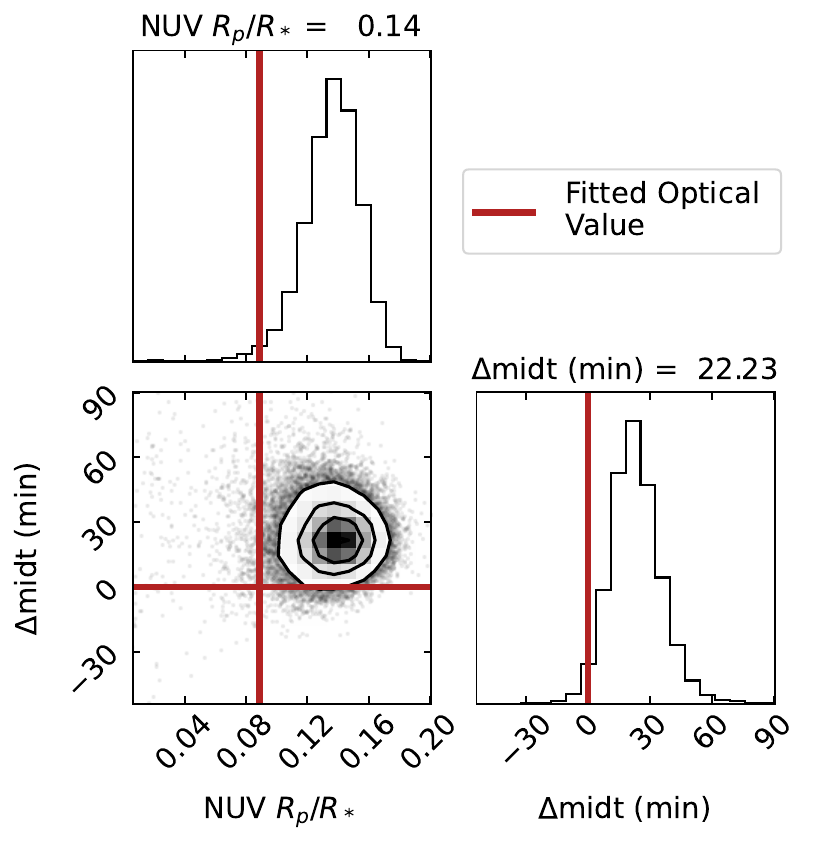}
    \caption{(Left) Posterior distribution for the optical transit fit parameters, including both the ground-based ISAS data and TESS data. The optical limb darkening coefficients were also a free parameter in this fit, but are not shown (see Table~\ref{tab:results}). The solid lines indicate past reported values for the impact parameter from \citet{Patel&Espinoza2022}, and the $R_p/R_{\star}$ and orbital period from \citep{Kokori2023}. (Right) Posterior distribution for the NUV transit parameters, fit simultaneously with the optical light curves. The solid lines mark the corresponding best-fit values for the optical $R_p/R_{\star}$ and expected mid-transit time from the fitted optical ephemeris, demonstrating that the NUV transit displays a larger radius and later central time. 
    }
    \label{fig:Corners}
\end{figure}

Our analysis 
also provides an updated solution for orbital parameters. We compare our posterior results to results from previous studies, namely the $R_{p,{\rm opt}}/R_{\star}$ from \citet{Patel&Espinoza2022} and the impact parameter and orbital period from \citet{Kokori2023}, in the left panel of Figure \ref{fig:Corners}. In the right panel, we compare the posterior distributions for the NUV radius, optical radius, and ephemeris.

\section{X-ray properties of XO-3}
\label{sec:Escape}

The X-ray luminosity of XO-3 provides a benchmark for estimating the atmospheric escape rate of XO-3~b. 
Simultaneous with the NUV observations, XMM-Newton monitored XO-3 with the three European Photon Imaging Cameras (EPIC). The Solar particle background did not flare at any point in the observations, enabling us to use the full exposure. XO-3 was clearly detected in all three cameras, providing the first X-ray characterization of this star. 
We examined the $0.3-2.4$~keV light curve, finding that the average count rate was approximately 9 ct~s$^{-1}$ and that variation in the light curve was negligible within the uncertainties. 
We thus treated the full observation as quiescent, and extracted a single X-ray spectrum for each of the three cameras in the $0.3-2.4$~keV band, using a minimum of 10 counts per bin.

We fitted a single temperature \texttt{apec} model \citep{Smith2001} to each of the spectra, using \texttt{Xspec}~12.14.0h \citep{Arnaud1996}. We included a \texttt{tbabs} term to account for interstellar absorption \citep{Wilms2000}. We estimated the line-of-sight ISM column using the \texttt{LISM\_NHI} code \citep{Youngblood2025, LISMcode}, which  
estimates $N_{\rm HI}$ absorption columns to a maximum distance of 100~pc. Since XO-3 lies at $205.4\pm1.1$~pc, we doubled the estimated hydrogen column to a value of $N_{H} = 10^{19}$~cm$^{-2}$. 
Abundances were fixed to Solar \citep{Asplund2009}, and we used C-statistics \citep{Cash1979} for fitting the models. 
The spectra from EPIC-pn and EPIC-MOS were fit simultaneously with the same model.

Table~\ref{tab:xrayResults} shows the best-fit parameters. We list two unabsorbed X-ray fluxes at Earth: one for the directly observed and fitted $0.3-2.4$~keV band, $F_{\rm X,0.3}$, and one for the commonly used $0.1-2.4$~keV ROSAT band, $F_{\rm X,0.1}$, calculated by extrapolating our best-fit model to softer energies. To estimate the EUV luminosity, we extrapolated from the $0.1-2.4$~keV band using the approach of \citet{King2018}. 

Our X-ray flux measurement is somewhat higher than the \textit{Swift} XRT upper limit reported by \citet{Corrales2021}, calculated by converting the measured counts into a flux upper limit. Using a more rigorous statistical method based on \citet{Kraft1991}, we calculated a 95\% confidence upper limit on the \textit{Swift} observations of $4.1\times 10^{-14}$\,erg\,s$^{-1}$\,cm$^{-2}$. This value assumes the same plasma temperature as the \textit{XMM-Newton} best-fit, and $N_{H} = 10^{19}$. Thus there is no statistical evidence of variation in the X-ray emission of XO-3 between the observational epochs of \textit{Swift} and \textit{XMM-Newton}.

Our measured 0.1--2.4\,keV X-ray luminosity of $\left(4.91^{+0.41}_{-0.42}\right) \times 10^{28}$\,erg\,s$^{-1}$ is towards the lower end of that observed for typical F stars \citep[e.g.][]{Topka1982,Micela1996,Pizzocaro2019,Johnstone2021}. Comparing with Figure~2 of \citet{Shimura2025}, our best-fit temperature and emission measure are lower than most F stars in their sample, but the combination of the two matches the few F stars in this region of the plot well. Since coronal temperature decreases with age, this may suggest that XO-3 is somewhat older than many F stars whose X-ray emission has been measured previously.


\begin{deluxetable}{lcc}
\tablecaption{Best fitting X-ray spectral fitting parameters and associated unabsorbed fluxes and luminosities.}
\label{tab:xrayResults}
\tablehead{\colhead{Parameter} & \colhead{Value} & \colhead{Unit}}
\startdata
Temperature, $kT$ & $0.336^{+0.025}_{-0.021}$ & keV   \\
Emission Measure, $\rm EM$ & $\left(2.33^{+0.22}_{-0.21}\right) \times 10^{51}$ & cm$^{-3}$         \\
X-ray flux at Earth, $F_{\rm X,0.3}$    & $\left(7.67^{+0.54}_{-0.61}\right) \times 10^{-15}$                     & erg\,s$^{-1}$\,cm$^{-2}$ \\
X-ray lum., $L_{\rm X,0.3}$    & $\left(3.87^{+0.28}_{-0.31}\right) \times 10^{28}$                      & erg\,s$^{-1}$            \\
X-ray flux at Earth, $F_{\rm X,0.1}$    & $\left(9.72^{+0.82}_{-0.83}\right) \times 10^{-15}$ & erg\,s$^{-1}$\,cm$^{-2}$ \\
X-ray Lum., $L_{\rm X,0.1}$    & $\left(4.91^{+0.41}_{-0.42}\right) \times 10^{28}$  & erg\,s$^{-1}$            \\
XUV Lum., $L_{\rm XUV}$        & $\left(1.45\pm0.20\right) \times 10^{29}$   & erg\,s$^{-1}$\\
XUV flux at planet, $F_{\rm XUV,p}$ & $21000\pm3000$ & erg\,s$^{-1}$\,cm$^{-2}$ \\
\enddata
\tablecomments{The subscripts $_{\rm 0.3}$ and $_{\rm 0.1}$ refer to the $0.3-2.4$~keV and $0.1-2.4$~keV energy bands, respectively. The XUV flux at the planet is time-averaged across the orbit, with the extrapolation to the EUV based on \citet{King2018}}.
\end{deluxetable}

\subsection{Mass loss rate of XO-3~b}

Given the non-zero orbital eccentricity of XO-3~b, we used the time-averaged separation between the planet and star \citep{Williams2003} to calculate the average XUV flux at the planet from the $L_{XUV}$ listed in Table~\ref{tab:xrayResults}. Therefore the mass loss rates we calculated are an average across each orbit.
Using energy-limited escape with canonical literature assumptions of a 15\% efficiency and the photons being absorbed at a negligible height above the measured planetary radius \citep[$\beta=1$ in equation 2 of][]{King2024}, we calculated a  mass loss rate of $5.5\times 10^9$\,g\,s$^{-1}$. This is 1-2 orders of magnitude lower than the mass loss rate of prototypical hot Jupiter planets with directly observed atmospheric escape \citep[e.g.][]{Etangs2010, Etangs2004, Lammer2003}.

The energy-limited escape rate is likely an over-estimate because XO-3~b is a particularly massive exoplanet (around 12~M$_J$) and so possesses a deep gravitational well that will 
make it more difficult for material to be lofted beyond the large Roche lobe radius of the planet ($\sim 10$~R$_{J}$). 
We employed the \texttt{ATES} analytical approximation to calculate an alternative effective escape efficiency which encompasses the absorption radius \citep{Caldiroli2022}, and should be more realistic for an XO-3~b-like planet than the energy-limited assumptions. This approximation is based on earlier hydrodynamic simulations by the same authors \citep{Caldiroli2021}, and results in a very low effective efficiency of $3\times10^{-7}$. 
This implies a very tenuous escape rate of just 12,000\,g\,s$^{-1}$. Such a low mass loss rate would suggests that the NUV absorbing material could come from an 
extended atmosphere that is still gravitationally bound, or otherwise arising from a different physical mechanism. 

\section{Discussion} \label{sec:Discussion}

\subsection{Comparison to \textit{Swift} transit data}
\label{sec:SwiftCompare}

\citet{Corrales2021} report an apparent NUV radius of $R_{p,NUV}/R_{\star} = 0.18 \pm 0.01$ for XO-3~b, with a transit midpoint that was early by $22.2 \pm 4.8$~minutes. At the time of publication, this was within 2$\sigma$ of the propagated ephemeris uncertainties of the best-fit planetary orbit. In this work, we provide a more precise orbital solution that does not give strong evidence for optical TTVs. Additionally, the TESS orbital solution is used to constrain the NUV transit light curve in this work, whereas the \textit{Swift} study held many of the orbital parameters fixed. This warrants revisiting the archival \textit{Swift} dataset.

Two transits were observed by \textit{Swift} on July 14th and October 5th of 2019 with durations of 3.364 hours. These data are plotted in Figures 2 and 8 of \citet{Corrales2021}. We re-reduced the UVOT dataset and extracted light curves using Heasoft v6.35.1. 
Similar to Section~\ref{sec:NUV}, we fitted the \textit{Swift} transits jointly with the TESS lightcurves used in this work, using GP and the limb-darkening parameters from the first column of Table~\ref{tab:results}\footnote{This was done with the \texttt{Juliet} transit fitting software \citep{Espinoza2019}. The limb-darkening curve parameters were fixed to the same values as used in this work and the same 3/2 Celerite GP kernel was used. Sampling was performed using the \texttt{Dynesty} sampler \citep{Speagle2020}.}. The two \textit{Swift} transit visits were not phase-folded, as was done by \citet{Corrales2021}, and were treated independently.

We performed four different fitting procedures to explore the robustness of the reported \textit{Swift} transit solutions. One did not allow for timing offsets while the other three allowed for a $\Delta t_{mid, s1}$ and $\Delta t_{mid, s2}$ transit midpoint offset in transit visit 1 and 2, respectively. 
Table \ref{tab:swiftfit} gives the adopted priors and the posteriors obtained for these two parameters, as well as for R$_{\rm{p}}$/R$_{\star}$ (which had a uniform prior between 0 and 0.5) seen by \textit{Swift} across the different fits. The `Fixed' model forced the \textit{Swift} transit midpoints to match the optical orbital solution from TESS. 
The `Loose' model was the most flexible, allowing the transit midpoint to vary across different transit visits and in any direction (early or late). We used a normal distribution prior centered on 0 with a standard deviation of 43.2~minutes. We chose this broad standard deviation to provide a sufficient range so that the model could fit a negligible or significant 
transit offset.
In this case, both of the transits appear early but still consistent with no transit offset. To see how well a late or early transit could be constrained (if at all), we performed the 'Late' and 'Early' fits with priors for both $\Delta t_{mid}$ values fixed to +30~minutes and -30~minutes, respectively, and with tighter variance compared to the loose parameter. 
The Bayesian Information Criterion (BIC) calculated for each fit favors a model where there are no transit offsets. The resulting posterior is almost identical the prior distribution, showing that the \textit{Swift} transit light curves---treating each visit separately and fitting simultaneously with TESS---are insufficient to constrain any offsets between the NUV and optical transit midpoints.

Despite these limitations, the \textit{Swift} data is always well fit with a transit depth consistent with \citep{Corrales2021}, $R_{p,NUV}/R_{\star} \approx 0.18$. We note that these these transits are also detected to approximately $3\sigma$ significance, except when we apply a prior that encourages a late transit. The range of best-fit $R_{p,NUV}/R_{\star}$ values and size of the error bars ($\approx 0.05$) demonstrate clearly that the error bar of 0.01 from \citep{Corrales2021} was underestimated. This is likely because, in that work, all XO-3~b orbital parameters were fixed to literature values, and thus did not incorporate the uncertainty from the orbital solution. In this work, we fit for the XO-3~b orbital parameters alongside the NUV transit data, while incorporating ultra-precise TESS light curves. These new constraints make our measurement with XMM-Newton the most robust and precise determination of the XO-3~b  $R_{p,NUV}/R_{\star}$ to date, with error bars that are three times smaller  than what could be accomplished with the \textit{Swift} dataset.

\begin{deluxetable}{lcccccc}
\tablecaption{\textit{Swift} transit fit constrained by TESS}
\label{tab:swiftfit}
\tablehead{\colhead{} & \colhead{Fixed} & \colhead{Loose} & \colhead{Late} & \colhead{Early}}
\startdata
Prior $\Delta t_{mid}$ (min) & 0.0 & $\mathcal{N}(0,43.2)$ & $\mathcal{N}(30,15)$ & $\mathcal{N}(-30,15)$ \\
Posterior $\Delta t_{mid, s1}$ (min) & 0.0 & $-28.80^{+25.92}_{-20.16}$ & $28.80\pm14.40$ & $-36.00\pm11.52$ \\
Posterior $\Delta t_{mid, s2}$ (min) & 0.0 & $-15.84\pm23.04$ & $23.04\pm14.40$ & $-28.80\pm12.96$ \\
Posterior R$_{\rm{p,NUV}}$/R$_{\star}$  & $0.17^{+0.04}_{-0.05}$  & $0.19^{+0.04}_{-0.05}$  & $0.12^{+0.06}_{-0.07}$   & $0.20 \pm0.04$ \\
$\Delta$BIC & 0 & +22 & +16 & +24  \\
\enddata
\tablecomments{The same $\Delta t_{mid}$ prior distribution was applied to $\Delta t_{mid, s1}$ and $\Delta t_{mid, s2}$. The $\mathcal{N}$ symbol indicates a normal distribution. The change in BIC was computed relative to the `Fixed' model; a positive value indicates a worse fit.}
\end{deluxetable}

\begin{figure}
    \centering
    \includegraphics[scale=0.45]{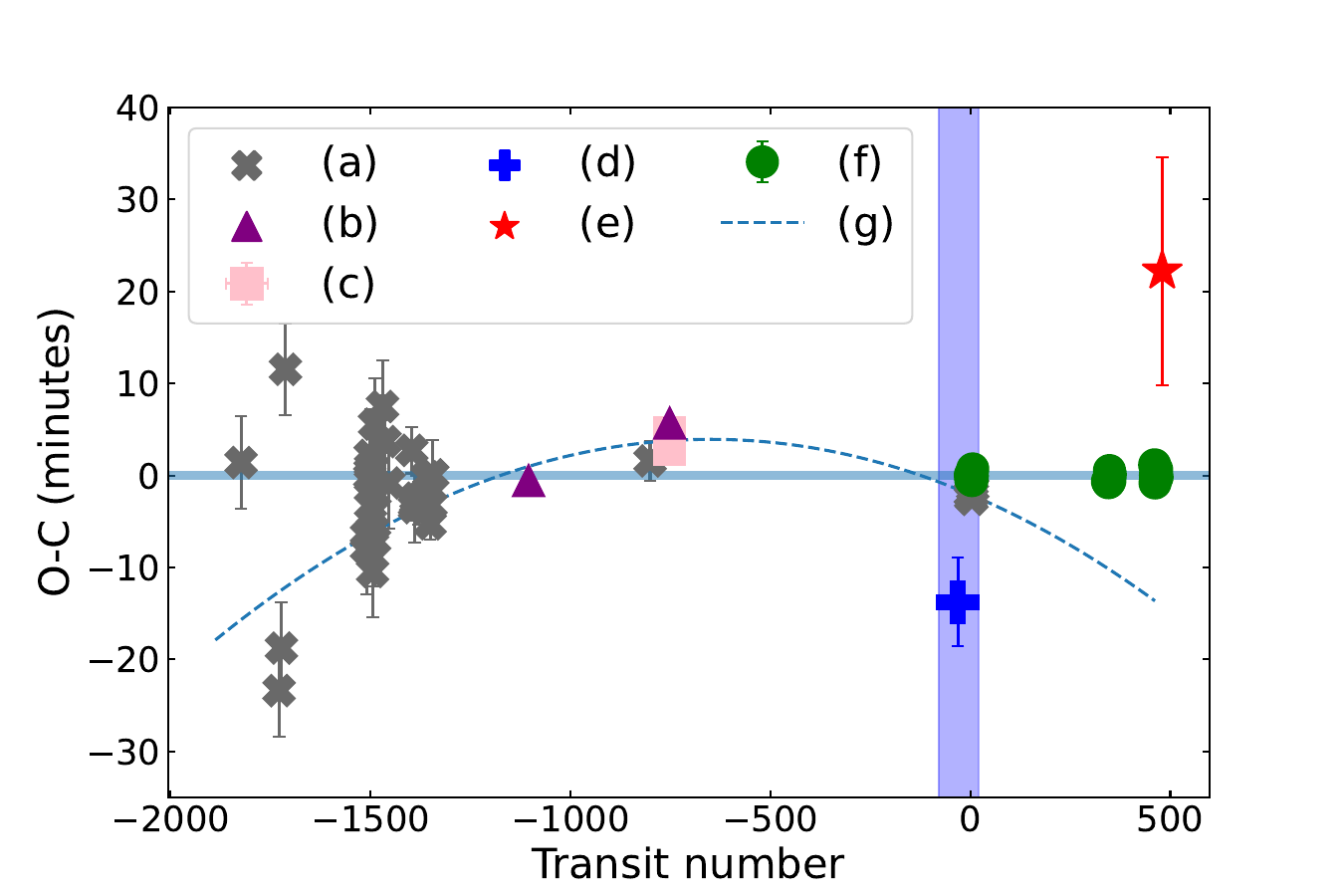}

    \caption{Second panel of Figure 2, now incorporating the NUV transit data. The gray points (a) are archival mid-transit times reported in \citet{Ivshina2022}. The purple (b) and pink (c) points are Spitzer observations from \citet{Wong2014} and \citet{Dang2022} respectively. The blue point (d) shows the \textit{Swift} transit observed by \citet{Corrales2021}, and the vertical blue bar represents the confidence interval from our reanalysis of that dataset. 
    The red point (e) comes from the XMM-Newton OM NUV transit analyzed in this study, and the green points (f) are the TESS transits analyzed in this study. The blue dashed curve (g) is the period-decay model reported by \citet{Yang2022}. It is evident that the XMM-Newton OM transit does not match the expected optical transit timing, whether or not the orbital period is considered to be constant.}
    \label{fig:NUVOmC}
\end{figure}

\subsection{Possible implications of an NUV transit offset}
\label{sec:NUVtiming}

In Figure \ref{fig:NUVOmC}, we add to the transit timing comparison plot of Figure \ref{fig:TESSTiming} the XMM-Newton OM NUV transit, as well as the reported early \textit{Swift} NUV transit from \citet{Corrales2021}. 
The temporal proximity of the NUV transit observations with the TESS data set, for which no TTVs are found, suggests that the late NUV transit we measure is not 
due to orbital variations. 
In this section, we investigate alternative explanations of the late NUV transit. 

The timing of an NUV transit relative to the optical can reveal important physics about star-planet interactions. One source of potential NUV absorption is Fe{\sc ii} in an extended, escaping atmosphere \citep[e.g.,][]{Sing2019}. Escaping atmospheres can sometimes produce cometary tails \citep[e.g.,][]{Ehrenreich2015}, in which case we could expect the asymmetries in the NUV transit shape that skews the apparent mid-transit times to be later than the optical transit.

We searched for asymmetries in the NUV transit light curve by mirroring the pre-transit data about the best-fit mid-transit time. We removed the first 14 minutes of ingress data, to match the end of the transit egress that was not included in the observation. 
We then performed a Kolmogorov-Smirnov (KS) test to compare the in-transit points before and after the NUV mid-transit time. The KS test yielded a KS statistic of 0.23 and a p-value of 0.32, which indicates that the mirrored pre-transit and post-transit datapoints are not likely to originate from different distributions. This means that the shape of the portion of the NUV transit included in the observation is statistically consistent with being symmetric. However, 
the missing $\sim$14 minutes of the egress, along with the post-transit baseline, could contain asymmetries that are not represented in this dataset, and we cannot completely rule out an asymmetric feature like a cometary tail as a cause of the apparent late transit. More observations of the entire NUV transit are necessary to draw conclusions about its symmetry.

An offset in the observed NUV mid-transit time might also be explained by a bow-shock induced by the interaction of the planetary magnetic field with the stellar environment. These bow shocks can cause an NUV transit to be either early or late depending on the orbital speed of the planet relative to the stellar plasma \citep{Vidotto2010, Vidotto2011a}. Furthermore, for eccentric planets, the orientation of these bow-shocks can vary across the orbit or with asymmetries in the stellar magnetic fields and winds \citep{Vidotto2011b}. There are two possible scenarios which could lead to a bow-shock: first, the planet could be passing through the escaping stellar wind, and second: the planet could be passing through coronal plasma that is confined by the stellar magnetic field and co-rotating with the star.
In both cases, the apparent NUV radius of the planet serves as an estimate for the size of the planetary magnetosphere, which grows with $B^{1/3}_p$, where $B_p$ is the planetary magnetic field strength \citep{Vidotto2011b}. Using an assumed $B_p$ and the true anomaly of the planet at transit (102.8°), we can estimate the temporal offset between the start of the optical and the NUV transit \citep[$\Delta t_{mid}$,][]{Vidotto2011b}.

XO-3~b is estimated to have, as an upper limit, a magnetic field of 250~G (20 times stronger than Jupiter), based on the proposed scaling relation for mean magnetic fields of rapidly rotating objects \citep{Christensen2009}, which was expanded to hot Jupiters while incorporating corrections for the incident flux on the planet \citep{Rakesh2017}. Taking the limits of possible planetary magnetic field strengths from 1 to 250~G in the stellar winds scenario, the $\Delta t_{mid}$ offset is 1.8~min ($B_p = 1$~G) to 55.2~min ($B_p = 250$~G). In the confined plasma scenario, the timing offset depends on the magnetic field strength of the host star. We assume a stellar magnetic field strength of 2~G, which is a typical expected value for an F-type star \citep{Seach2020}. Again assuming the limits of $B_p=1$~G to 250~G, the $\Delta t_{mid}$ offset for the confined plasma scenario is 0.01~min ($B_p = 1$~G) to 43.6~min ($B_p = 250$~G). Both of these estimates assume a bow-shock oriented at an angle of 90~degrees relative to the planet-star sight line, which means they represent an upper limit. Nonetheless, the suspected NUV transit timing offsets for XO-3~b sit comfortably within this range of possibilities.

Following the methods laid out in \cite{Vidotto2011b}, we calculated the conditions needed for a bow-shock to form around XO-3~b. A bow-shock will form when the orbital motion of the planet is supersonic, and thus host stars with colder plasma temperatures are more likely to produce this phenomenon. The elliptical orbit of XO-3~b causes it to experience varying plasma conditions. At the true anomaly of the transit, a bow shock can form for temperatures less than 
$2\times10^5$ K in the confined plasma scenario, and $2\times10^7$ K in the stellar winds scenario. The temperature of the X-ray emitting plasma of XO-3 ($4\times10^6$~K; Table~\ref{tab:xrayResults}) suggests that a shock could not occur in the confined plasma scenario.

To further test the stellar winds scenario, we calculated the isothermal stellar wind speeds of XO-3 using the \texttt{ParkerSolarWind} package \citep{Badman2025}. We assume the density at the base of the corona to be $1.26\times10^9$ cm$^{-3}$ following the method of \cite{Marin2017}. The Parker wind model predicts the density of the plasma during the transit
to be $7\times10^6$ cm$^{-3}$, which satisfies the minimal condition of $n > 10^4$ cm$^{-3}$ to produce observable absorption, as put forward by \cite{Vidotto2011a}. However, the stellar winds scenario cannot produce a trailing magnetospheric bow-shock. The coronal confinement scenario can produce a trailing bow-shock; however, for the true anomaly of the XO-3~b transit, we find that a leading bow-shock is expected. We are therefore unable to explain the late NUV transit using this simplistic model. More work, such as advanced magnetohydrodynamic simulations, are needed to assess whether or not XO-3~b can form a trailing NUV absorbing region as a result of star-planet interactions \citep[e.g.,][]{Duann2025}.

\section{Conclusion} \label{sec:Conclusion}

XO-3~b is a unique high-mass hot Jupiter in an eccentric orbit that leads to variable heating of its atmosphere, providing a unique regime to probe NUV absorption from atmospheric escape, aerosol formation or destruction, and magnetospheric interactions with a stellar wind. Our observations of the NUV transit via XMM-Newton demonstrate an anomalously large effective radius, $R_{p,NUV}/R_{\star} = 0.1371^{+0.016}_{-0.019}$ (implying 
$R_{p,NUV} = 2.05^{+0.28}_{-0.31} R_{J}$). This effective NUV radius is $30-70\%$ larger than the optical radius $R_{p, opt}/R_{\star} = 0.08868^{+0.0009}_{-0.0007}$, measured via 18 optical transit light curves.

An offset between the NUV and optical transit midpoint for XO-3~b was first reported by \citet{Corrales2021}. However, the ephemeris uncertainty at the time and later reports of TTVs for this system \citep{Yang2022} demonstrate the need to better constrain the optical ephemeris. We analyzed all available TESS light curves for XO-3~b, spanning 2019 to 2024, and found no evidence for TTVs within this epoch. We present new constraints on the XO-3~b $R_{p,opt}/R_{\star}$, orbital period, and impact parameter ($b$) that are distinct from previous literature values (Table~\ref{tab:results} and Figure~\ref{fig:Corners}). We also fit the TESS dataset alongside the NUV transit and a ground-based optical light curve obtained simultaneously with the XMM-Newton observations. The high precision of the TESS light curves provide strong constraints on the orbital period and ephemeris for the epoch of the NUV light curve, which was obtained in 2024. The apparent $22^{+13}_{-11}$~minute-long late offset of the NUV transit midpoint could provide clues to the physical origin of the NUV absorption.

The apparent late transit of XO-3~b in 2024 contradicts the reported early transit from the \textit{Swift} dataset obtained in 2019. We reanalyzed the two \textit{Swift} transit light curves, without phase-folding, simultaneously with the TESS dataset. The best quality fit was obtained when the \textit{Swift} transit midpoint was fixed to the predicted optical ephemeris. When we fitted for the transit midpoints of each \textit{Swift} light curve as free parameters, no strong constraint could be made. Additionally, by allowing for more free parameters in the transit model and anchoring with TESS data, we demonstrated that the original error bars for the \textit{Swift} NUV transit were significantly overestimated. Nonetheless, the reanalyzed \textit{Swift} transit depth is consistent with the new observations by XMM-Newton, which are more precise with $\sim 14\%$ error.

Simultaneous with the XMM-OM, the XMM-Newton EPIC and MOS X-ray detectors allowed us to study the X-ray properties of the host star XO-3 for the first time. No flares were detected over the course of the transit, and the quiescent $0.3-2.4$~keV X-ray luminosity was $\left(3.87^{+0.28}_{-0.31}\right) \times 10^{28}$~erg~s$^{-1}$ at the time of the transit; consistent with the upper limit X-ray flux we derived from \textit{Swift} data. We used this value to estimate the average XUV flux on XO-3~b and the level of XUV-driven escape. Because XO-3~b is so massive ($\gtrsim 11$~M$_J$), it should hold on to a H/He atmosphere. Standard energy-limited analytic estimates suggest a mass loss rate of $5 .5\times 10^{9}$~g~s$^{-1}$. However, the more complex hydrodynamic prescriptions from ATES \citep{Caldiroli2022} predict a miniscule escape rate of $\sim 10^4$~g~s$^{-1}$. These estimates bode poorly for atmospheric escape as an explanation for the extended NUV absorption by XO-3~b.

An overdensity of material in a bow-shock built up by a magnetospheric interaction with stellar wind plasma might also lead to enhanced NUV absorption. This mechanism could lead to NUV transit offsets on the order of a few to 10s of minutes compared to the optical. Using simple analytic prescriptions set forward by \citet{Vidotto2011a}, it is possible for XO-3~b, which could have a magnetic field strength on the order of 100~G, to produce an NUV absorbing bow-shock. However, this simple analytic description does not predict a late transit as suggested by the XMM-OM dataset. 

Future work is needed to fully investigate the physical nature of XO-3~b's NUV absorption. A major source of uncertainty for our conclusions is the lack of a full egress and post-egress baseline measurement, preventing more concrete characterization of any stellar activity or asymmetries in the transit shape. Unfortunately, XO-3 is at a limited viewing angle for XMM-Newton, observable only in segments of 30-40~ks. The data presented in this work indeed takes up the entire observability window for XO-3. More detailed computational studies are also needed to investigate and predict the NUV transit depth and patterns. Simulations of planetary magnetospheric interactions with a stellar magnetosphere and plasma environment are much richer than the analytic prescriptions investigated here \citep{Vidotto2025}. Furthermore, computer simulations are needed to explore the circulation patterns and heating that can lead to aerosol formation and destruction patterns -- which could produce SiO signatures -- in the unique environment of this eccentric hot Jupiter.

\section{Acknowledgements}


\software{\texttt{Lightkurve} \citep{Lightkurve}, \texttt{Juliet} and its dependencies \citep{Kreidberg2015,Espinoza2019,Speagle2020}, \texttt{exoplanet} and its dependencies \citep{exoplanet:joss,
exoplanet:zenodo, celerite1, exoplanet:agol20, exoplanet:arviz,
exoplanet:astropy13, exoplanet:astropy18, exoplanet:luger18}}

This work has made use of data from the European Space Agency (ESA) mission
{\it Gaia} (\url{https://www.cosmos.esa.int/gaia}), processed by the {\it Gaia}
Data Processing and Analysis Consortium (DPAC,
\url{https://www.cosmos.esa.int/web/gaia/dpac/consortium}). Funding for the DPAC
has been provided by national institutions, in particular the institutions
participating in the {\it Gaia} Multilateral Agreement.

All TESS data used can be found in the MAST archive at the following DOI: \dataset[10.17909/v28p-z130]{http://dx.doi.org/10.17909/v28p-z130}.

This research is made possible by the support of NASA Grant 80NSSC23K1580 through the  XMM General Observer (GO) Program. Portions of this research are also supported by the Cottrell Scholar Award CS-CSA-2024-068 sponsored by Research Corporation for Science Advancement.

\bibliography{sample631}{} 
\label{sec:bib}
\bibliographystyle{aasjournal}

\end{document}